\documentclass[iop]{emulateapj}
\usepackage{amsmath}
\usepackage{natbib}
\usepackage{color}
\usepackage[normalem]{ulem}

\catcode`_=\active
\newcommand_[1]{\ensuremath{\sb{\mathrm{#1}}}}
\newcommand{\be}[1]{\begin{equation} \label{eq:#1}}
\newcommand{\ee}{\end{equation}}
\newcommand{\ba}[1]{\begin{eqnarray} \label{eq:#1}}
\newcommand{\ea}{\end{eqnarray}}

\newcommand{\pref}{\protect\ref}
\newcommand{\iris}{{\em IRIS}}

\newcommand{\solrad}{\ifmmode{R}_{\rm S}\else${R}_{\rm S}$\fi}
\newcommand{\solmas}{\ifmmode{M}_{\rm S}\else${M}_{\rm S}$\fi}

\newcommand{\tintu}{\ifmmode{\rm erg~cm^{-2}~s^{-1}sr^{-1}}\else 
  erg~cm$^{-2}$~s$^{-1}$~sr$^{-1}$\fi}
\newcommand{\fluxu}{\ifmmode{\rm erg~cm^{-2}~s^{-1}}\else 
  erg~cm$^{-2}$~s$^{-1}$\fi}
\newcommand{\velu}{$\,$km$\,$s$^{-1}$}

\newcommand{\wave}{\ifmmode{\lambda} \else$\lambda$\fi}

\newcommand{\hot}{\ifmmode{8\times10^4~{\rm K}}\else{$8\times10^4$~K}\fi}

\newcommand\lta { \mathrel {\hbox to 0pt {\lower 3.7pt \hbox{$\sim$}
      \hss} \raise 1.7pt \hbox{$<$}}}
\newcommand\gta { \mathrel {\hbox to 0pt {\lower 3.7pt \hbox{$\sim$}
      \hss} \raise 1.7pt \hbox{$>$}}}

\newcommand{\new}[1]{#1}

\newcommand{\philemail}{judge@ucar.edu}
\newcommand{\alinaemail}{alina.donea@monash.edu}
\newcommand{\danielaemail}{daniela.lacatus@monash.edu}
\newcommand{\alinemail}{alin.paraschiv@monash.edu}
\newcommand{\charlieemail}{lindsey@cora.nwra.com}

\newcommand{\figbefore}{
\begin{figure*}[h] 
\epsscale{0.8}
\plotone{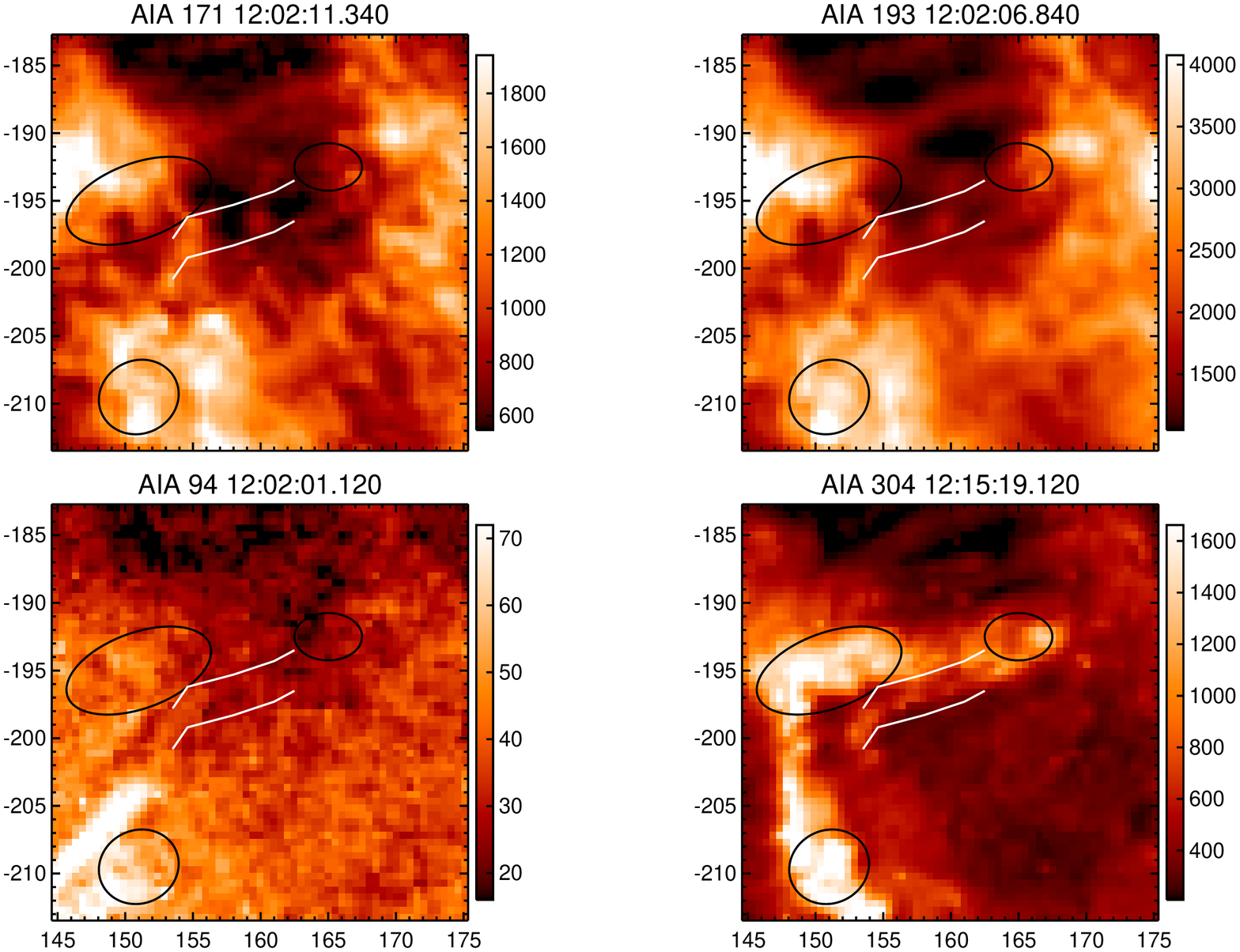}  
\caption{\label{fig:before} {Intensity data from the AIA
    instrument on the SDO spacecraft are shown as a function of
    heliographic solar $X$ and solar $Y$ coordinates, in seconds of arc.  The field of
    view spans part of active region 11890 on 2013 November 9th.  The
    data reflect the chromospheric and coronal environment
    surrounding the flare ribbon under examination. In AIA data, the
    ribbon is seen most clearly, at 12:15:19 UT, in the 304 \AA{} band
    (predominantly \ion{He}{2}) in the bottom right panel. All other panels show conditions {\em prior to the ribbon's appearance}, close to 12:02 UT. The ribbon
    lies between the two white lines.  Black ellipses identify regions
    where, during the brightening at 12:15:19 UT, the 304 \AA{} emission
    is associated with moss emission and vertical magnetic field
    (see Figure~\pref{fig:hinode}). 
    Typical moss emission is seen all around the ribbon in the bright
    patches of network plage emission, for example at $X=170,Y=-190$,
    The color bars label instrument DN values.  } }
\end{figure*}
}

\newcommand{\figone}{
\begin{figure*}[h] 
\epsscale{1.15}
\plotone{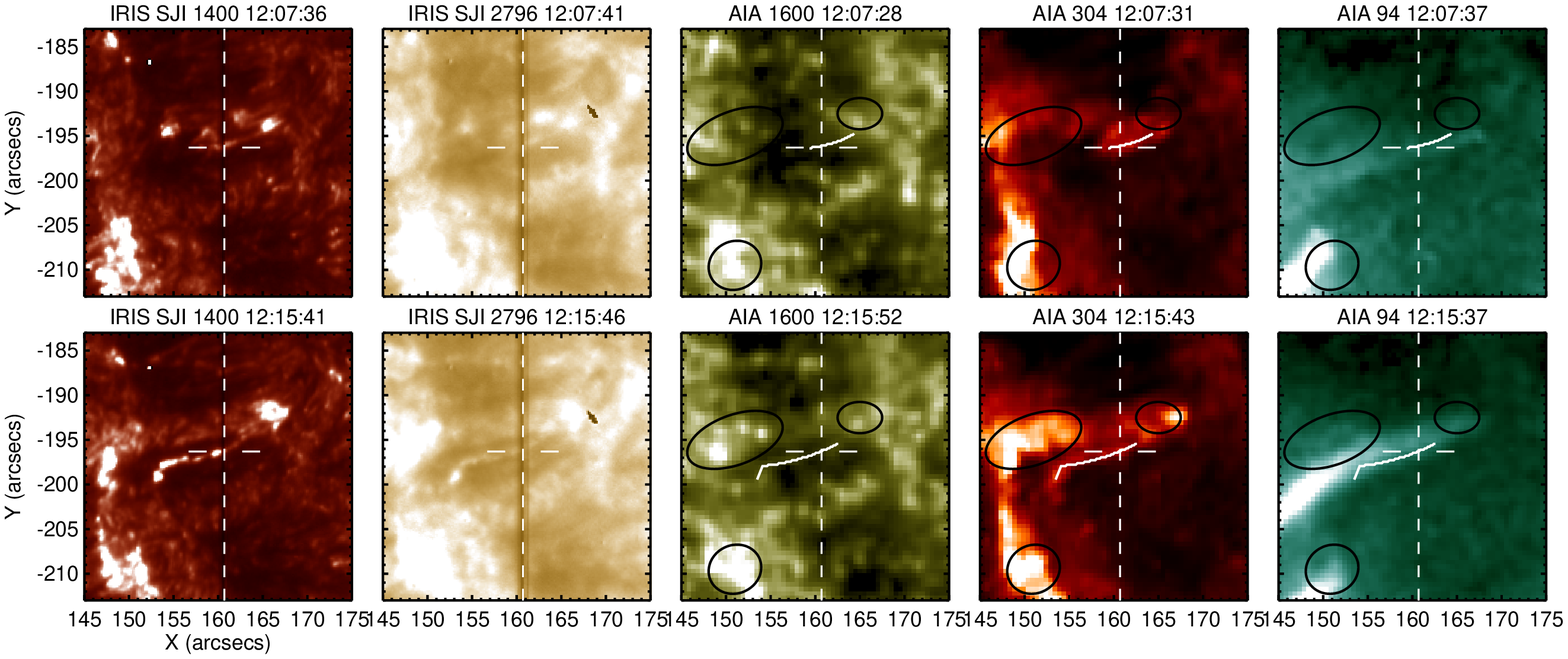}  
\caption{\label{fig:only} Intensity data are shown for  active region 11890 on 2013 November 9th, corresponding to two {small} flaring events associated to the ROI. Upper panels show the weaker 12:07 flare ribbon analyzed by \citet{Testa+others2014}. Lower panels  show the  stronger flaring peaking at 12:15.  All panels have been co-aligned by eye to approximately $0\farcs3$. The solar coordinates are referred to the coordinate system specified in the headers.  Only a $30\arcsec \times 35\arcsec$ field of view is shown, to highlight the precise position of the flare ribbon traced in all panels except for the slit-jaw \iris{} panels.  {As before, point ``D''
of Figure 2 in T2014 is marked by the intersection of the vertical dashed line with the 
short horizontal marked line.}
}
\end{figure*}
}

\newcommand{\figtwo}{
\begin{figure*}[h] 
\epsscale{1.15}
\plotone{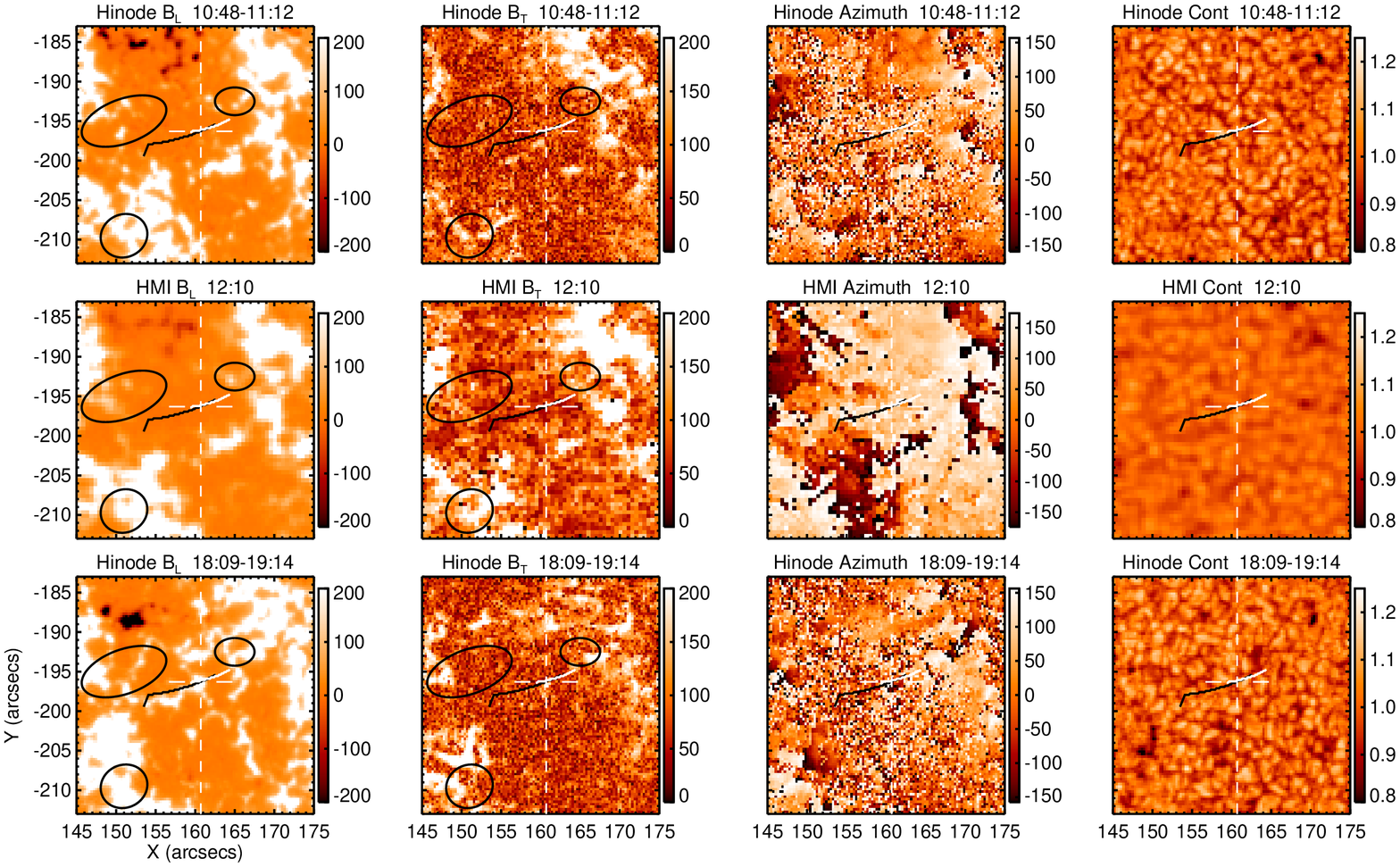}  
\caption{\label{fig:hinode}Magnetic  data are shown for
  active region 11890, before, during, and after the two flares at 12:07 (ribbon extent traced with a white line) and 12:15 (black line) shown in
Figure~\pref{fig:only}.  Each row shows longitudinal, tangential fields, field azimuth (with the 180$^\circ$ ambiguity unresolved) and the continuum intensity.  The upper panels show the Hinode scan 
obtained before the flares, the middle panels show HMI data closest to the flaring events, and the lower panels show Hinode data from the next complete scan some six hours 
later.  Six hours is a considerable fraction of the lifetimes of
supergranules, so surface field evolution has occurred between the two HINODE
scans. High cadence HMI magnetic field evolution animation for the time-range in between the two Hinode scans is available as electronic supplementary material. }
\end{figure*}
}

\shortauthors{P. G. Judge,  A. Paraschiv, D. Lacatus, A. Donea, C. Lindsey}
\shorttitle{Flare ribbons}

\slugcomment{}

\begin{document}

%###############################################################################
%
%     OPENING
%
%###############################################################################

\title{Are all flare ribbons simply connected to the corona?}
\author{  
Philip G. Judge }
\affil{High Altitude Observatory,\\
       National Center for Atmospheric Research\thanks{The National %
       Center for Atmospheric Research is sponsored by the %
       National Science Foundation},\\
       P.O.~Box 3000, Boulder CO~80307-3000, USA; \philemail}

\author{Alin Paraschiv, Daniela Lacatus,  } \affil{
 Monash Center for Astrophysics, School of Mathematical Science,\\
  Monash University, Victoria 3800, Australia; \\ and Institute of Geodynamics ``Sabba S. Stefanescu" of the Romanian Academy, \\ 19-21, str. J.L. Calderon, Bucharest, Romania;\\ \color{black}
  \alinemail, \danielaemail } 
  
\author{Alina Donea } \affil{Monash Center
  for Astrophysics,\\ School of Mathematical Science,
  Monash University,\\ Victoria 3800, Australia; \alinaemail}
  
\and

\author{Charlie Lindsey}
\affil{Northwest Research Associates, Boulder, CO, USA
\charlieemail}

%###############################################################################
%
%     ABSTRACT
%
%###############################################################################

\begin{abstract}
We consider the observational basis for the belief that flare ribbons in the chromosphere result from energy transport from the overlying corona.  We study ribbons of {small} flares using magnetic as well as intensity data from the Hinode, SDO and \iris{} missions.  While most ribbons appear connected to the corona, 
{and they over-lie regions of significant vertical magnetic field,}
we examine one ribbon with no {\new clear} evidence for such connections.  Evolving horizontal magnetic fields seen with Hinode suggest that reconnection with pre-existing fields {below the corona } can explain the data.  The identification of just one, albeit small, 
ribbon, with no apparent connection to the corona, leads us to 
conclude that at least two mechanisms are responsible for the heating that leads to flare ribbon emission. 
\end{abstract}


\section{Introduction}

The purpose of the present paper is to test the hypothesis that flare
ribbons are always associated with the downward transport of energy
released in the corona.  These ribbons are locations where the much 
 of the flare  energy is radiated back into space.
{Specifically, we will analyze a ribbon reported as being 
caused by ``nano-flaring'' activity, motivated by the 
peculiar environment in which this ribbon seems to exist, and noting that  the 
presence of ribbons is characteristic of flares
of all sizes. Figure~\pref{fig:before} shows the environment prior to the 
brightening of the  ribbon under study, as well as the ribbon itself some minutes later, as seen in the 304 \AA{} channel of the 
using  data from the AIA instrument 
\citep{Lemen+others2012} 
on the Solar
Dynamics Observatory (SDO).}

\figbefore

{Ribbons are 
taken to be one signature of a universal process involving 
magnetic reconnection.}   In this ``standard model'' of flares
\citep{Carmichael1964,Sturrock1966,Hirayama1974,Kopp+Pneuman1976},
non-thermal energy assumed to be stored 
for some time in the corona is suddenly released
\citep{Gold+Hoyle1960}. 
The energy
propagates away from the 
release site in the corona. When propagating downwards, bright ribbons and
kernels of radiation are observed as this directed energy is dissipated 
in dense chromospheric plasma.  

The basic hypothesis of the standard model is supported by a considerable wealth of 
evidence \citep[see][for a recent review]{Fletcher+others2011}.
The advent of routine hard X-ray (HXR) 
data simultaneously with soft X-rays and EUV/UV data
from modern spacecraft 
has confirmed that HXR emission arises from Bremsstrahlung
radiation emitted as energetic electrons 
impact the denser chromosphere \citep{Brown1971}.  The connection between
energetic electron precipitation and flare ribbons has become generally accepted.   
 
A particularly clear example of the magnetic corona-chromosphere connection is seen 
in Figure 3 of \citet{Sadykov+others2016}. 
Yet there are indications that the relationship between energy
propagating down from the corona and that observed coming from the
denser layers is not quite this straightforward.  For
example, flare ribbons are bright in regions where 
 hard X-rays are undetectable as well as bright
\citep[e.g., section 3.3 of][]{Fletcher+others2011}.  A clear example 
of a bright ribbon covering a more extended area than HXR emission is
shown in Figure~1 of \citet{Judge+Kleint+Sainz-Dalda2015}.  Quoting from
\citet{Fletcher+others2011}:
\begin{quote}
``In general the HXR sources are confined to localized areas situated
  on the outer edges of the elongated flare ribbons observed in UV and
  H$\alpha$ and are predominantly associated with bright H$\alpha$/UV
  kernels\ldots''
\end{quote}
What, then, is the relationship between beams of particles, and more
generally, energy transport, from the corona to the flare ribbons?
The question is of some importance, for if energy is stored and
released in regions other than the corona, then the mechanism for
flare energy storage and release is not restricted to the corona,
i.e. the standard model does not describe all flares. 

Recently \citet[][henceforth ``T2014'']{Testa+others2014}.   used \iris{} data \citep{dePontieu+others2014}
to infer the presence of
non-thermal particles by combining data from \iris{} with synthetic data.  The synthetic data are required because an instrument like \iris{} cannot by itself be used to infer non-thermal particles. 
T2014 conclude that 
\begin{quote}
``The observable that discriminates most efficiently between the beam
  heating models and the conduction models is the Doppler shift in the Si IV emission.''
\end{quote}
We note that the post-impulsive 
dynamics as reflected 
in \ion{Si}{4} or other UV emission is far removed from the energy transport and dissipation mechanisms. Thus we ask if 
other scenarios for energy storage and release might be acceptable. 

%\figmotives

We decided to re-examine the data examined by T2014. For most of their
ribbons, magnetic data (and ``moss" emission) support the idea that a
coronal magnetic connection exists prior to the ribbon emission (see the ellipses in Figure~\pref{fig:before}).
However the ribbon highlighted by white lines 
in this Figure has no clear  connections to the
overlying corona, and that appears to be {\em incompatible with the
  standard model}. {\new
This particular ribbon drew our attention when comparing the image
of 1400\AA{} with that of 193\AA{} in Fig. 1 of T2014, taken during the
outburst. We noticed the dramatic differences in the relative
brightness of the ribbon (1400/193 \AA{} brightnesses) compared with the
surrounding network footpoints. The latter are identified as those regions where the
accompanying ``hot'' 94\AA{} loops end. We noted that the ribbon of interest is far more
intense relative to the coronal emission during the flare.  Our Figure
1 shows data acquired before the flare, also showing that the
overlying ``moss" emission (patchy emission at 171 and 193 A) has a
different morphology, with essentially zero emission seen
across half of the area later covered by the ribbon.}

{It is these differences that lead us to ask if all flare ribbons 
result from energy transport down from the corona.  For without our careful
observations, one is led to conclude that the ribbon of interest is merely 
yet another manifestation of the processes captured by the standard model.  
T2014 modeled part of this ribbon {\em assuming} the standard model. }
Indeed both power law distributions and arguments concerning the
universal nature of reconnection in space physics support the notion
that flare physics might be independent of scale
\citep[e.g.][]{Nishizuka+others2009,Shibata+Takasao2016,Sharma+others2016}.
{Our work should be viewed as a test of the general validity of the standard model
as applied to solar flare-ribbon phenomena. }

\section{Observational analysis}

\subsection{UV and EUV data}

We refer to T2014 for the circumstances of the \iris{} and other observations of active region 11890 on 2013 November 9th.   
Figure~\pref{fig:only} shows images of interest from \iris{} and the AIA instrument.  The ribbon of central interest here contains the point ``D'' analyzed by T2014,  lying $252\arcsec$
SW of Sun center.  The local vertical is  $\vartheta=15^\circ$ from
the line-of sight, with $\mu=\cos\vartheta= 0.96$. {We will refer to this ribbon as the 
``ribbon of interest'' (henceforth ``ROI'')}.

During the \iris{} observations the chromospheric ribbon brightened twice, of which only the first, weaker, event was discussed by T2014. Here we concentrate on the second brightening which is similar in morphology. This makes no difference to our conclusions because we examine the magnetic connectivity between the corona and ribbon prior to both brightenings.  The bright ribbon is clearly seen at its brightest phase (12:15:41 UT) in the lower left panel of Figure~\pref{fig:only}.  {This 
ribbon's locus is traced in the other images using white lines. 

\figone

The ribbon is {barely seen} in the 1600 \AA{} AIA
channel, which admits contributions from the UV continuum, the
resonance lines of \ion{C}{4} and the
Balmer-$\alpha$ line of \ion{He}{2} (1640 \AA).
It is clearly seen in the  
\ion{Mg}{2} slit-jaw image 
(Figure~\pref{fig:only}) which forms across the entire 
chromosphere. 
The AIA image at 94 \AA{} appears superficially 
to overlap with
the ribbon. However in movies S1 and S2 published
by T2014, the 94 \AA{} emission clearly 
originates from closer to the larger UV flaring
patch seen near $(X,Y) =(167,-192)$, {highlighted with 
a black ellipse  in our figures.}
The 94\AA{} coronal emission appears close to but over the ribbon of interest in a loop-like structure connected to the region of 150 G line of sight field, near $(167,-192)$.   
We have examined all other channels from the AIA
instrument between 12:15 and 12:16 UT.  The
clearest indicator of the ROI seen in AIA data is in the
304 \AA{} channel (mostly \ion{He}{2}).  It is not
detected in the 1700 \AA{} UV continuum channel,
but during the ribbon's brightening 
there are hints of weak emission at 171, 193,
211 and 335 \AA{}, all coronal bands.  

In conclusion, the near-absence of 1600 \AA{} ribbon emission and its presence 
in lines from \ion{Mg}{2} to \ion{He}{2} 
indicate that the ROI's emission originates from the 
mid-to-upper chromosphere and the transition region.}

\subsection{Magnetic data}

Figure~\pref{fig:hinode} shows magnetic parameters inverted  from raster scans with the SP instrument on the Hinode spacecraft \citep{Lites+others2008}. We used data from the CSAC at HAO based on the code MERLIN.  The coordinates are all rotated to a reference frame close in time to the two ribbon flaring events obtained using the HMI instrument \citep{Schou+others2012} on SDO. But, given the {small} size of the ribbons, we chose to base our further analysis on the Hinode raster scans due to the higher spatial resolution and magnetic sensitivity. In spite of the evolution of solar features during the Hinode scans (the 30$\arcsec$ widths of the images shown {took} $\approx 6$ minutes, comparable to a granule lifetime), the Hinode images shown have been successfully co-aligned to those of HMI to $\approx \pm 0\farcs3$.  
The magnetic data are striking. There is
essentially just noise in the line-of-sight (longitudinal) component
of the field $B_L$ near the {ROI}.  The Hinode data give a mean value of
$11\pm10$ G for the longitudinal field calculated from areas extending 2$\arcsec$ around the ribbon loci shown.  The noise in longitudinal field acquired with the SP  corresponds to just 3 Mx~cm$^{-2}$ in each spatial pixel \citep{Lites+others2008}.  The average Hinode tangential (near-horizontal) field component $B_T$ is $73 \pm 24$ G.

If similar connections existed from the ROI it would have several such origins distributed along its locus in the 94 \AA{} or other coronal images.

\figtwo

The raster scans from the Hinode
spectropolarimeter shown in Figure~\pref{fig:hinode}  were obtained about 
1 hour before and 6 hours after the
flaring activity.  An intermediate scan was started but failed during the flaring period, it contains no useful data. But from the data shown, 
three characteristics caught our attention.  Firstly,
the 200G horizontal flux seen near the ROI before the flaring is
replaced by 200G vertical field components hours after the flare.
Secondly, the azimuth of the field in the neighborhood of the ROI
before the flare has significant 
spatial coherence, with a mean value close to 
${44^{\circ}}_{-27}^{+31}$ relative to the N-S direction.  This aspect suggests, albeit weakly, the 
presence of an emerging magnetic field roughly along the same direction as the long axis of the ribbon.  An HMI time-series animation of the field evolution in between the two Hinode raster scans is provided as supplementary material.  HMI unfortunately does not have the sensitivity to reveal horizontal fields in the upper photosphere of the ROI during this period. 

\section{Discussion}

T2014
recognized that chromospheric reconnection might explain some of their data:

\begin{quote}
``Chromospheric reconnection could in principle provide an
  alternative explanation for the observed chromospheric and TR
  variability... The moss brightenings {\em clearly occur at
  conjugate footpoints of hot loops undergoing heating}, and there is a
  {\em clear correlation between the coronal and chromospheric/TR emission},
  naturally explained by beam heating.'' 
\end{quote}

Above, we have added italics to those statements that appear to be
inconsistent with the ROI.  It is important to
note that all other bright ribbons and kernels lie above or close to regions with 
detectable longitudinal fields.   The absence of
a longitudinal field above about $10$ G in the photosphere near the ROI suggests that
there is little magnetic flux emerging through this region into the corona. 

\subsection{The magnetic field prior to the flaring of the ROI}

The question of connection or the ribbon to the overlying corona concerns only the connection between chromosphere and corona (such small flares generate undetectable changes in deeper layers). The magnetic field threading the chromosphere must spans several scale heights and must therefore be
significantly different than measured in the photosphere, which therefore presents a difficulty. But given the measured photospheric fields it is hard to see from where a chromospheric field connected upwards might originate.  The
entire stratified chromosphere spans about 1500 km, which corresponds to  $\approx 2\arcsec$.  In Figure~\pref{fig:hinode} the ribbon is
encased in a void of field with values $\lta10$ G extending at least $2\arcsec$ all
around the ribbon.  We already noted 
the absence of moss emission in this area (Figure~1 of T2014).

Some weak longitudinal field may suffice to provide unseen connections to the corona along the ribbon.
Indeed field is present to the south of both ends of the ROI, of the same polarity, 30-40 G in 
strength and covering just a couple of square arcseconds (see the
Hinode panel $B_L$ in the top left of Figure~\pref{fig:hinode}). 
These fields are not visible in HMI by the time of the flare, 1 h 22 m later  (second row of the Figure).  
These small concentrations are some $5\arcsec$ from the center of the ROI.  Could 
longitudinal fields such as these have been advected to the locus of the ROI  by known surface motions in time for a coronal connection to be established? \citet{Berger+others1998} studied surface diffusion rates around active regions, finding
with a coefficient of $\kappa\approx$ 80 km$^2$~s$^{-1}$ for several tens of minutes.  
In the $t=100$ minutes after the first Hinode scan, elements would diffuse a distance $\frac{1}{3} \sqrt{\kappa t}$  $=\frac{1}{3}\sqrt{80\times6000} \approx 200$ km $0\farcs3$ using this coefficient. Supergranular diffusion in regions far from
strong fields is larger, with 
$\kappa\lta$ 800 km$^2$~s$^{-1}$. Even with such a high rate of diffusion, the surface diffusion would  have to accumulate from within a distance of 1$\arcsec$ of the ribbon, and produce vertical components along the ROI
in the beam heating scenario.  This seems  unlikely given the quasi-random nature of this diffusion implicit in the work of \citet{Berger+others1998}. 

\subsection{The magnetic field after the disappearance of the ROI}

In the 6 hours between the flare and the second  Hinode scan) relatively strong post-flare longitudinal``network" fields appear in the lowest panel of 
Figure~\pref{fig:hinode}.  It is seen meandering underneath the position of the ROI, with a length $\ell \approx 10\arcsec$. 
Either emergence and/or surface diffusion of field must be responsible for this new network flux. 
The new network flux amounts to $ \Phi \approx 200$G times the apparent area of the network, $\approx 1\arcsec\times\ell\arcsec$ $\equiv 5\times10^{16}$ cm$^{2}$, so that $\Phi \approx 10^{19}$ Mx. To build this flux from the mean pre-ribbon longitudinal field strength of $10$G,  this flux would have to have collected, merged and strengthened from an area of about 200 square arcseconds, about 100 Mm$^2$.  This area is far larger than the area available for diffusion in the time available, which we write as 
$L\ell \approx \frac{1}{3} \kappa 21600$ $\approx 6$ Mm$^2$, even using $\kappa \approx 800$ km$^2$s$^{-1}$.  

It seems that some flux emergence into the chromosphere occurred between the first and second Hinode scans. Such flux emergence is below the detection limit for HMI (row 2, column 2 of Figure \pref{fig:hinode}) but is hinted at by the coherent horizontal field seen in the first Hinode scan (row 1, column 2 of Figure \pref{fig:hinode}).  We suggest that this flux emergence might help explain the presence of the ROI, in particular the horizontal fields that appear somewhat close to 
the ribbon in the pre-ribbon Hinode scan strongly indicate horizontal flux.  The series of HMI vector field maps was examined between 09:58 and 20:10 UT, from 11 to 19 UT is shown in the supplememntal movie.   From this movie, any horizontal fields are too noisy to be seen as any coherent structure rising through the photosphere.  Beginning close to 18:00 UT, the longitudinal field patch close to (-165,-201) migrated quickly towards the position of the clump of flux seen underlying the ROI in the last Hinode (18:09-19:14 UT) scan. 

Even if we cannot conclusively detect horizontal emerging flux, we can conclude that the {\em magnetic conditions around the ROI are qualitatively different to the other ribbons discussed by T2014}.

\subsection{Energy requirements and the coronal magnetic field}

But can we discount a coronal origin for the ROI?
Let us consider the energy requirements, for this we can use figure 3 of
T2014.   From this figure we see
that the \ion{Mg}{2} $k$ line has excess emission
averaging 400 DN/second/spatial pixel for a period of 200 seconds. Using calibration factors {\em SolarSoft} routine {\tt iris\_get\_response.pro}, we find the
intensity of \ion{Mg}{2} $k$ is $I_k = 1.5\times
10^5$ \tintu{}.
The $k$ line contributes $\approx 10\%$ of the total chromospheric radiative
losses known in 1981  \citep[Table 29
  of][]{Vernazza+Avrett+Loeser1981}, and allowing
for a doubling of the losses due to the inclusion
of \ion{Fe}{2} lines \citep{Anderson+Athay1989}, we find
that the excess emission is $\approx 20 \times
4\pi I_k \approx 4\times 10^7$
erg~cm$^{-2}$~s$^{-1}$.  The same calculation for
the \ion{Si}{4} 1403 \AA{} line gives an excess 
intensity
of $6\times10^3$ \tintu, 25 times smaller than
seen in the $k$ line.

We 
assume that we have a field strength
of 10 G, corresponding to the $3\sigma$ upper
limit, emerging from the region of the ROI
into the corona.  (Below we relax this assumption).  
The total energy per unit area
in a straight tube of length $L$ is $L B^2/8\pi$,
which for $L \approx 10^{10}$ cm (100 Mm, as used
by T2014) and $B\lta 10$ G 
gives $\lta 4\times 10^{10}$ erg~cm$^{-2}$.  The
304 \AA{} frame in movie ``S2'' of
T2014 shows that the heating of
the ROI continues beyond the time span of the \iris{} sequence
which ends at 12:17:16 UT. Thus we can set a time
$\gta 10^2$ seconds for the duration of the
heating.  Therefore the overlying corona can
supply $\lta 4\times 10^8$ erg~cm$^{-2}$~s$^{-1}$ of magnetic power per
unit area into the chromosphere below.  If,
optimistically, we assume that 10\% of the
magnetic energy along this tube is free energy
available, then just $\lta 4\times 10^7$ erg~cm$^{-2}$~s$^{-1}$
is available.  This is close to the energy
requirements of the quiet Sun chromosphere
\citep{Anderson+Athay1989b} and our 
excess estimate above.  An alternative 
limit can be derived using the Alfv\'en crossing time 
for the tube is $L/V_A$. With a pre-flare coronal 
gas pressure 
$p \approx 0.1 $ dyne~cm$^{-2}$ \citep{Jordan1992}, $T=10^6$ K, we
have a density $\rho \approx 2\times 10^{-15}$ g~cm$^{-3}$ and 
 $V_A \lta 700$ \velu. Then $L/V_A \gta 140$ sec.  

The energy available in a coronal tube is formally
enough to account for the radiation losses from
the chromosphere below.  But our estimate is a strict
upper limit because: 1. We have used the $3\sigma$
``detection'' of longitudinal field from the
Hinode scan 1 hour before the flare, the energy is
$\propto B_L^2$. The field strength in the overlying corona 
is surely lower than the 10 G upper limit in the photosphere;
 2. The energy is 
$\propto L$, and $L$ has been taken from neighboring 
coronal structures seen clearly in the 94 \AA{} channel
that are rooted in far stronger concentrations; 3. We have assumed 
that 10\% of {\em all} the magnetic energy is available for 
release within the tube, and 4. We have assumed that sufficient time
exists for this energy to be released. Reconnection typically 
takes place at $0.1V_A$, through complex dynamical mechanisms
such as the plasmoid instability \citep[e.g.][]{Bhattacharjee+others2009}. 
Thus 
the dynamical time for energy release 
becomes too long $L/0.1V_A \approx 1400$ sec. 

It appears that the ROI cannot be reconciled by heating 
from the overlying corona.  Either the dynamical time is too long
($L= 100$ Mm) or, if $L$ is small enough to account for the dynamical
time, then the free energy $\propto L$ is insufficient to account for the 
observed losses. 

Now let us permit the coronal connection to
include the {\em transverse} component of the
field as well as the longitudinal component.  Then
we expect the connection to the corona to be
highly oblique to the line of sight, as is seen in the 94
\AA{} image (Figure~\pref{fig:only}).  In this case the
energy available is some fifty times bigger, and
the Alfv\'en speeds seven times larger.  In this
case, the observations are reconciled with a beam
heating picture: the energy release rate from the
corona is $\lta 2\times 10^9$ \fluxu{} and the dynamical time is
about 200 sec, compared with about $2\times 10^7$
\fluxu{} and 100 sec respectively.

But is this scenario credible?  The morphology of the horizontal field and the absence of moss  coronal emission simply does not support the idea that the fields detected are connected to the corona. 

\subsection{Local heating}

We are forced to consider {\em local} sources of
heating for the ROI.  The tangential field in
the Hinode scan is around 75 G.  Let us consider
the emergence of a near-horizontal tube of flux
through the chromosphere in a quasi-steady state
prior to some energy release process.  It emerges
into pre-existing field in the upper chromosphere
in the active region where a pre-existing canopy
field exists, fields rooted elsewhere in stronger
concentrations which have turned horizontal to
ensure pressure balance as the chromspheric gas
pressure drops well below the magnetic pressure. 
The total
energy density available in an emerging 
tube with this field strength 
is $B_T^2/8\pi \approx
220$ erg~cm$^{-3}$.  The Alfv\'en speed
is roughly $B_T/ \sqrt{4\pi \rho_u} \approx 200$
\velu{} where $\rho_u \approx  10^{-12}$
g~cm$^{-3}$ is the density in the upper 
chromosphere.  Let us assume that the tube 
emerges into canopy fields, rooted in more distant network patches, of a lower strength. To estimate
this strength we note that in the neighborhood of the ribbon
the longitudinal flux is all of the same sign, 
average longitudinal flux density over a region centered at 
the ribbon is about $55-80$ Mx~cm$^{-2}$, depending on the local
area chosen.  For the surrounding $20\arcsec\times20\arcsec$ area (of order 
one supergranule in size) the average is 80 Mx~cm$^{-2}$. 
We envisage that the locally emerging flux begins to 
reconnect with the canopy field and generate the extra emission
in the ROI.  To account for the ribbon, we require, as above,
$4\times10^7$ \fluxu{} for about 200 seconds.  
Once the reconnection starts, it will proceed at a rate close to
$0.1 V'_A$, where $V'_A$ is the Alfv\'en speed in 
the reconnecting component of the field.  
If the angle between the 
emerging tube and the canopy fields is $\vartheta$, then  
$V'_A  = V_A \sin \vartheta$ 
Thus the reconnection will convert magnetic into thermal
energy at the rate 
\be{recon}
{\cal F} \lta 0.1 \sin^3 \vartheta 
\frac {B_T}{\sqrt{4\pi\rho_u}}
\frac {B_T^{2}}{8\pi} 
\lta 5\times10^7 \fluxu.
\ee
where we have used $\sin \vartheta=0.5$ as a rough
estimate of the mismatch in directions of the
horizontal and canopy fields.  It seems that there
is power available from chromospheric
reconnection to drive the radiation losses seen in
the ROI.  With a (vertical) inflow speed into
the reconnection layer of $0.1 \sin\vartheta V_A \approx 25$
\velu{}, lasting for $\approx 200$ seconds, the reconnection
advects a total mass per unit area of $m_R = 2.5\times10^6 \times 200
\rho_u = 5\times 10^{-4}$ g~cm$^{-2}$ from the chromosphere 
into the reconnection layer.  If this partially ionized plasma is
heated via dynamical instabilities (e.g., the plasmoid instability,
\citealp{Bhattacharjee+others2009})
or kinetic processes (ion-neutral collisions, for instance),
then this emerging flux effectively leads to heating in the upper chromosphere for column masses above 
$m_R=5\times10^{-4}$ g~cm$^{-2}$.  Interestingly, this corresponds roughly to
the range of heights between which the proposed beam dissipation arises 
in the models of T2014 (Fig.~S4).  Therefore, we would expect similar dynamical signatures from this type of heating as beam heating when the reconnected field has access to the overlying corona.

\subsection{Can we refute chromospheric reconnection for the ROI?}

One might argue that the chromosphere remains
``closed'' not open to the overlying corona, and
therefore that lines formed are trapped in closed
field lines under a low-$\beta$ regime in which
significant line shifts observed in \ion{Si}{4}
cannot be observed.  However, in this picture we
envisage plasma mass advected into the canopy
fields which are themselves open to the overlying
corona, a process commonly called ``interchange reconnection".   We would naturally expect the {\em same}
kind of dynamics along one of the canopy field
lines as computed in the beam heating scenario,
since both result from deposition of energy in
plasma of similar column masses, that later enters
via reconnection a tube of field that is connected
to the corona.  This ambiguity reminds us of the
sobering reality that lines such as \ion{Si}{4}
are like a dog with two masters- both the corona
and chromosphere have significant effects on such
transition-region lines.  Also, it must be
remembered that \ion{Si}{4} intensities are a
factor of 25 weaker than the \ion{Mg}{2} lines.
Thus, these lines reflect only a small fraction of
the energy release and its properties in the
evolving atmosphere.
 
A common reason for discounting chromospheric
reconnection is that the ribbons ``light up''
between observations separated by tens of seconds,
and that there is nothing that can communicate one
part of the chromosphere to reconnect on these
time scales.  But this is not correct near active
regions with transverse fields of order 70 G.
The ROI has a horizontal extend of about 7 Mm
(Figure~\pref{fig:only}), so it would take just 14
seconds for magnetic perturbations, such as a
tearing mode, to propagate horizontally across the
surface. Further, an emerging 
flux rope defines a ribbon-line morphology as it interacts
with the overlying fields. 
Therefore we respectfully disagree with the dismissal of chromospheric reconnection discussed in section S3 of T2014, in the case of the ROI. 

Finally, a widely held belief that the chromosphere is highly dynamic
has arisen from studies seeking dynamical phenomena associated with
the chromosphere \citep[e.g.][]{dePontieu+others2007}.  The question
would then arise if the upper chromosphere would shred a rope of
horizontal flux before it could interact with pre-existing fibrils
extending across sueprgranular cells.  But this is a non-issue since
simple estimates of such things as spicules and related phenomena show
them to cover no more than 0.1\% of the solar surface area, and to originate 
at supergranular vertices. Spectral
observations of the solar disk 
with HRTS by \citet{Dere+others1983} show both linewidths
and shifts very rarely approach those of the dynamic type II spicules
reported by de Pontieu and colleagues.  It is also clear from
narrow-band imaging observations of spectral lines formed in the upper
chromosphere, such as the \ion{Ca}{2} infrared triplet lines, that 
fibril structures are stable on time sacels of hours or more
\citep[e.g.][]{Cauzzi+others2008}.

\section{Conclusions}

We have shown that at least one of low-energy
flare several ribbons analyzed by
T2014 is very probably
magnetically disconnected from the corona.
Sufficient energy exists in the emerging magnetic
field for local dissipation in the upper
chromosphere to account for the observed behavior.
Our analysis therefore refutes the statement
(T2014) for at least one flare
ribbon:
\begin{quote}
``Our analysis provides tight constraints on the properties of such
  electron beams and new diagnostics for their presence in the
  non-flaring corona.''
\end{quote}
Care is needed in the interpretation of data from
UV instruments like \iris{}, where the UV spectra
are the end step of a series of complex non-linear
phenomena including the elusive heating terms in
the energy equation, otherwise known as the
long-standing ``coronal heating problem''.  

In conclusion, there are at least two processes
that can lead to the enhanced radiation seen in
small flare ribbons, one of which does {\em not}
rely on energy transport from the overlying corona.

Lastly, we note that the ribbons are {small} events
compared with large flares. It remains to be seen if
bigger events can be found in which an energetic connection to the 
corona can be refuted. 

\acknowledgments

Hinode SOT/SP Inversions were conducted at NCAR under the framework of
the Community Spectro-polarimtetric Analysis Center (CSAC;
\url{http://www.csac.hao.ucar.edu/}). This work was carried out during
a visit to Monash University by PGJ and visits of DL, AP and AD to
HAO, supported by HAO visitor funds and the School of Mathematical
Science, Moansh University.   We are grateful to a patient referee, in particular for their 
help in clarifying the observational  motivation behind this work.

\references

\end{document}